\documentclass[11pt]{article}

\usepackage[final]{acl}

\usepackage{times}
\usepackage{svg}
\usepackage{array}
\usepackage{float} 
\usepackage{rotating}
\usepackage{amsmath}
\usepackage{latexsym}
\usepackage{booktabs}
\usepackage{tikz}
\usepackage{url}
\usepackage{stfloats}
\usepackage{xcolor}
\definecolor{citehl}{RGB}{255,247,230}  
\usetikzlibrary{arrows.meta,positioning,fit,calc,shapes.multipart,shapes.geometric}

\usepackage[T1]{fontenc}

\usepackage{graphicx}
\usepackage[dvipsnames]{xcolor}

\usepackage{listings,xcolor}
\lstdefinestyle{prompt}{
  basicstyle=\ttfamily\footnotesize, breaklines=true, frame=single,
  backgroundcolor=\color{gray!8}, columns=fullflexible, showstringspaces=false,
  xleftmargin=4pt, xrightmargin=4pt, aboveskip=4pt, belowskip=4pt}

\usepackage[utf8]{inputenc}

\usepackage{microtype}

\usepackage{inconsolata}

\usepackage{graphicx}
\usepackage[dvipsnames]{xcolor}
\title{Citing Less Critically: LLMs Reshape the Rhetoric and Reach of Scientific Citation}

\author{
 \textbf{Yixuan Liu\textsuperscript{*1}},
 \textbf{Lin Chen\textsuperscript{*1,2}},
 \textbf{Zhuoqi Liu}\textsuperscript{1},
 \textbf{Jianglin Lu}\textsuperscript{3},
 \textbf{Dakota Murray}\textsuperscript{4},
\\
 \textsuperscript{1}Network Science Institute, Northeastern University \\
  \textsuperscript{2}Department of Physics, Northeastern University \\
 \textsuperscript{3}Department of Electrical and Computer Engineering, Northeastern University \\
 \textsuperscript{4}Department of Information Sciences and Technology, \\ University at Albany, State University of New York
 \\
\small{*Equal contributions.}
 \\
 \small{
   \textbf{Correspondence:} \href{mailto:dsmurray@ualbany.edu}{dsmurray@albany.edu}
 }
}

\begin{document}
\maketitle

\begin{abstract}
Scientific citations carry rhetorical intent.
Scholars may cite prior work positively (supporting), negatively (contrasting), or neutrally (mentioning). 
As large language models (LLMs) increasingly assist scientific writing, whether they reproduce citations with the same rhetorical intent as humans remains unclear. 
We introduce a masked-citation task to compare human and LLM-generated citation behavior. 
For each citation context, an LLM generates a replacement citation sentence, producing a counterfactual corpus directly comparable to human citation. 
We analyze what, whom, and how models cite, using an LLM-as-a-judge to classify citation intent and a 20-million-edge coauthorship network to measure social distance between cited authors. 
Across six popular LLMs and 1,746 top NLP conference papers (63k+ contexts, 132k+ citations), three patterns emerge: 
(1) Compared with human citation, LLMs cite significantly less critically;
(2) LLMs over-cite popular and older papers, a tendency amplified for contrasting citations where human writing more often draws on recent, niche work;
(3) Whereas humans often cite within their close social network, especially for supporting citations, LLMs tend to draw on more socially distant authors. 
Together, these differences are double-edged: LLM citation reaches beyond a scholar's close collaborators while being less critical and amplifying visibility bias, reshaping the rhetoric and reach of scientific citation. \footnote{Code for reproducibility: \url{https://github.com/liu-yi-xuan/llm_citation_intent/}.}
\end{abstract}

\section{Introduction}
Citations have an important role in science, requiring discretion on the part of authors. 
The choice of what to cite reflects the author's background, inspirations, and how they view their work in relation to their discipline.
At the same time, this choice is shaped by social forces, such as reputation, recency, and an author's own social network~\citep{merton_matthew_1968,wallace_small_2012}.
Citations also carry rhetorical intent~\cite{jurgens_measuring_2018,Shu2026}.
While many citations are neutral (\emph{mentioning})~\citep{moravcsik_citation_1988}, authors may also choose to endorse the referenced work (\emph{supporting}), or else to dispute or contrast it with their own (\emph{contrasting})~\citep{chen_origin_2024,lamers_investigating_2021,catalini_incidence_2015}.
Citations shape the scientific literature even after publication.
Once published, a paper's citations provide context for readers, connecting the publication to related works that a reader may pursue~\citep{woo_2024_retractions}, and in aggregate, serve as indicators of impact for evaluating papers and authors~\citep{wilsdon_2025_metric,caon2020,Leydesdorff2016}. 

LLMs have seen rapid uptake in scientific workflows, including drafting related work sections and recommending references~\citep{hao_hlm-cite_2024,kusumegi_scientific_2025, liang_quantifying_2025, zhao_llm_2026,li_autosota_2026}. 
This shift has the potential to change the meaning and utility of citations.
Whereas citations have historically reflected human choice, they may now be generated automatically using LLMs, with an unknown degree of human oversight. 
It is important to understand the extent to which LLM-generated citations differ from those written by humans, in order to better understand the consequences of these tools on the scientific literature.  

Recent works on LLM-generated citations mainly focus on three lines of study. 
The first line audits \emph{reliability}, documenting that LLMs frequently fabricate or misattribute references \citep{press_citeme_2024, niimi_hallucinations_2025, linardon_influence_2025}. 
The second line studies \emph{citation selection} \citep{algaba_large_2025,algaba_how_2026}, finding that LLM-recommended citations broadly reflect human citation patterns but with a heightened bias toward impactful and small-team works.
The third line studies \emph{author-level disparities}, examining the demographic characteristics of authors in LLM-generated citations, such as their gender or race \citep{tian_who_2024, he_who_2025}. 
However, none of them examines the \textit{rhetorical role} a citation plays, i.e., whether it supports, contrasts, or merely mentions the cited work, nor how the choice of what to cite varies with that role.

In parallel, the field of Natural Language Processing has a long tradition of citation-intent and citation-polarity classification \citep{teufel_automatic_2006, jurgens_measuring_2018, cohan_structural_2019}.
The \textit{science of science} has meanwhile characterized human citing behavior through cumulative advantage and the Matthew effect \citep{merton_matthew_1968, price_general_1976}, citation aging \citep{chen_origin_2024}, co-authorship homophily \citep{newman_structure_2001, wallace_small_2012}, and social proximity~\cite{kozlowski_citation_2025}. 
These studies, however, are focused on citations generated by humans.
It remains unclear whether such tendencies and biases are also reflected in LLM-generated citations.

In this paper, we aim to understand the impact of LLM citations through the following questions: 
\begin{itemize}
    \item \textbf{RQ1}: Do LLMs reproduce the distribution of citation intents observed in human citations?
    \item \textbf{RQ2}: Does citation intent modulate the divergence between LLM and human citations in cited-paper attributes?
    \item \textbf{RQ3}: Does coauthorship proximity shape LLM citation behavior across intents, as it does for humans?
\end{itemize}

We address these research questions by constructing a corpus of aligned human- and LLM-generated citation contexts. 
Because the human's choice of citation is known, every LLM-generated citation serves as a counterfactual to the real human choice for the same position. 
We label each citation's intent via an LLM-as-a-judge approach and examine biases by intent, and then place its authors in their coauthorship network, measuring the citing--cited dyadic distance.

In contrast with earlier work documenting LLMs' heightened popularity bias, our design reveals a divergence in citation choice that is systematic across LLMs and conditioned on intent. 
Prior audits treat citation as a flat retrieval or recommendation under varied prompting setups~\citep{algaba_large_2025, algaba_how_2026, walters_fabrication_2023, tian_who_2024, he_who_2025}.
None that we know of preserves the local rhetorical context of each masked citation slot, conditions human--LLM comparisons on citation intent, or measures author-level coauthorship distance from focal to cited authors; the closest social-side precedent is a single aggregate self-citation rate \citep{algaba_how_2026}, whereas we compute network distances over a 20.3M-edge coauthorship network spanning 2.1M researchers. Our methodological and empirical contributions include:

\begin{itemize}
\itemsep0.2em
\item \textbf{A position-aligned, slot-level benchmark}: A position-matched evaluation framework in which LLMs recover each post cut-off masked citation in its rhetorical context, counterfactual to the human choice at the same slot.
\item \textbf{The first intent-conditioned comparison of LLM- and human-generated citations}, showing that LLMs under-produce contrasting citations and that popularity and recency biases are amplified along the intent dimension.
\item \textbf{An author-level socio-structural analysis} of the citing author's coauthorship network, showing that humans tend to cite papers within their close social neighborhood, especially for supporting citations, whereas LLMs cite more socially distant papers.
\item \textbf{LLM citation is double-edged}: it may broaden selection beyond a scholar's narrow social circle, but tends to cite older, more popular work in a less critical tone, motivating continued audits of LLM-assisted citation.
\end{itemize}


\section{Methodology}

\begin{figure*}[!tbp]
    \centering
    \includegraphics[width=\linewidth]{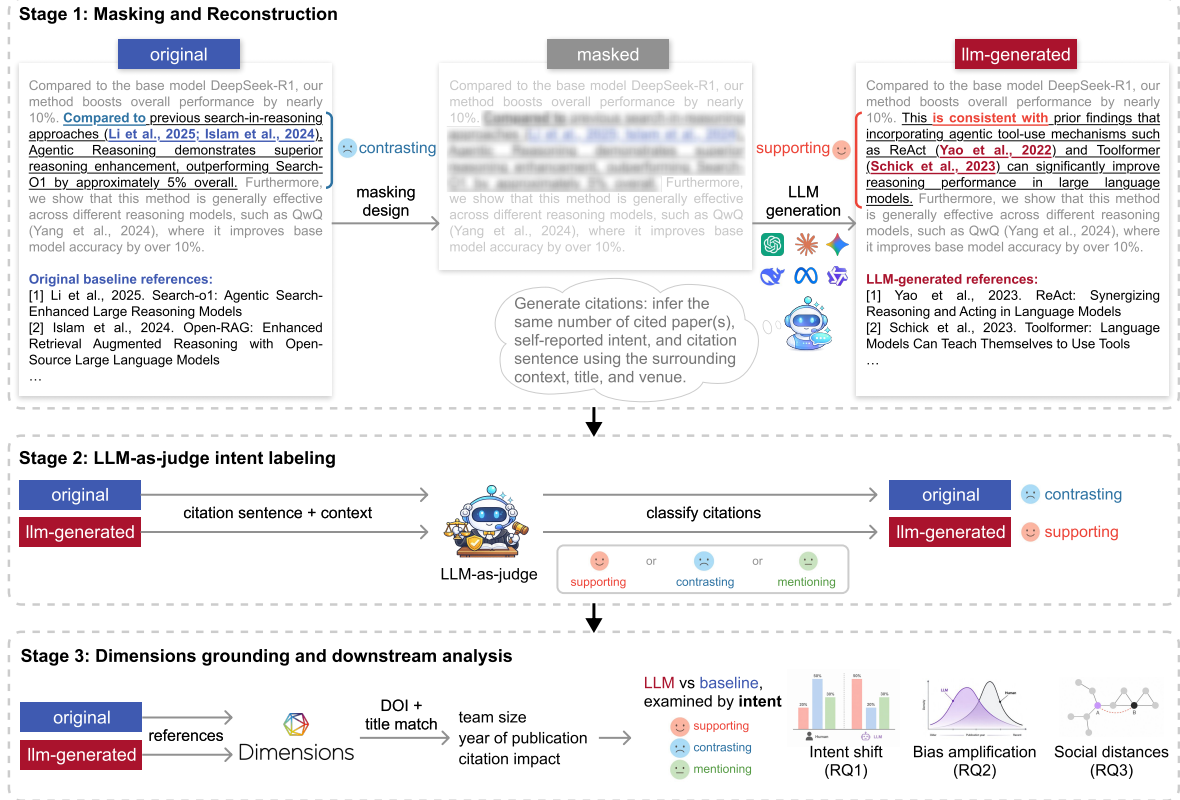}
\caption{\textbf{Framework for auditing how LLMs reshape scientific citation behavior by intent.}
From 1{,}746 main-track papers (with available \texttt{.tex} and \texttt{.bib} from arXiv), we extract every citation sentence and run the pipeline:
\textbf{Stage 1}: Masking and reconstruction. Each citation sentence is replaced by a masked placeholder, preserving title, venue, section, $\pm$1-sentence context, and required citation count. LLMs independently fill each slot with real papers, a fill sentence, and a self-reported intent.
\textbf{Stage 2}: LLM-as-judge intent labeling. An independent judge labels every original and reconstructed sentence as \emph{supporting} / \emph{contrasting} / \emph{mentioning} on identical context, without seeing cited-paper titles, so the label reflects the rhetorical move, not which work is cited. The example paper from ACL 2025 shows the intent shifting from human \emph{contrasting} to GPT-5.1 \emph{supporting}.
\textbf{Stage 3}: Dimensions grounding and downstream analysis. Recommendations are matched to Dimensions via DOI then normalized title, returning team size, recency, and citation counts that feed the downstream analyses: intent shift (RQ1), bias amplification (RQ2), and social distance (RQ3).}
    \label{fig:work-flow}
\end{figure*}

\subsection{Task definition}
We cast citation as a reconstruction task: for each citation context in a corpus, an LLM is presented with a version in which the citation sentence is masked with a placeholder (Figure~\ref{fig:work-flow}; example in Figure~\ref{fig:mask-example}, prompts in Figure~\ref{fig:prompt}). 
As input, the LLM receives the citation context, consisting of the one sentence before and one after the masked citation sentence.
Additionally, the LLM is provided with the heading of the section in which the citation context appears, the number of works originally cited in the sentence, and the citing paper's title, venue, and year of publication (see Appendix~\ref{app:context} on varying context windows).
The LLM is prompted to suggest the same number of real papers as originally cited and to generate a reconstructed sentence for the masked position.
For each recommendation, the LLM also provides a self-declared citation intent (supporting/contrasting/mentioning) and a confidence level.

Importantly, our design incorporates three constraints that prior work does not jointly satisfy: 
\emph{slot-level position matching} (every LLM citation is a counterfactual to the human choice at the same position, paired at the sentence level); 
\emph{post-cutoff evaluation} (all papers in our corpus were released as preprints after the six models' training cutoffs, preventing direct retrieval of the held-out citation); 
and \emph{count-controlled reconstruction} (the required number of citations matches the human baseline, so that tone/structure differences are not confounded by retrieval length).
Together, these constraints align the human and LLM outputs along position, time, and count, leaving citation choice as the remaining dimension of variation.

\subsection{Labeling citation intent}\label{main:label_intent}

\begin{table}[!htbp]
\centering\small
\begin{tabular}{p{0.21\linewidth} p{0.67\linewidth}}
\toprule
Intent & Definition \\
\midrule

Supporting
\includegraphics[width=0.37cm]{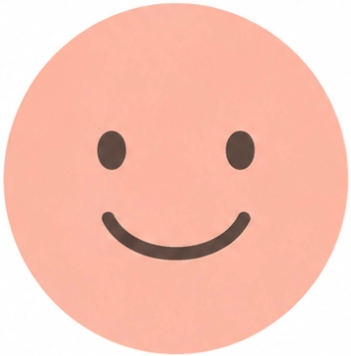} &
Cited work provides evidence, methods, or findings aligned with the citing paper's claims or approach. \\

Contrasting \includegraphics[width=0.37cm]{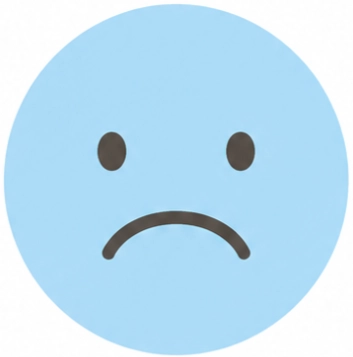} &
Cited work is a competing approach, contradicting finding, or baseline the citing paper improves upon or disagrees with. \\

Mentioning \includegraphics[width=0.37cm]{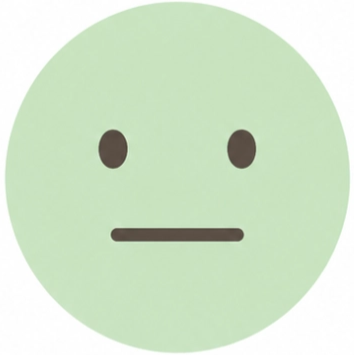} &
Cited work is referenced for background, definitions, or general acknowledgment with no clear support or contrast. \\

\bottomrule
\end{tabular}
\caption{Definitions of the three citation intents used throughout the study. Labels and definitions are inspired by \textit{Scite.ai} categories~\citep{nicholson_scite_2021}.}
\label{tab:motivation_definition}
\end{table}

To label citation intent, we use an LLM-as-judge approach: given a citation sentence and its context (title, section, adjacent sentences), the judge assigns an intent label (supporting / contrasting / mentioning, defined in Table~\ref{tab:motivation_definition}, illustration in Table~\ref{tab:motivation_definition_example}), and a confidence level (Figure~\ref{fig:work-flow}, example at Figure~\ref{fig:judge-example}, prompts in Figure~\ref{fig:judge-prompt}). 
The prompt provided to the LLM judge includes the citing paper's title, the heading of the section in which the sentence appears, and the citation context itself. This mirrors the standard setup in citation-intent classification, where intent is read from the citing sentence and its local context~\citep{jurgens_measuring_2018,cohan_structural_2019}.
The judge is not presented with information about the cited paper, instead inferring intent from the text alone. We further validate these labels against three human annotators, and the judge matches the human majority ($\kappa=0.60$), specifically for supporting and contrasting citation types (Appendix~\ref{app:human_val}).


\subsection{Matching citations to papers}
Each human- and LLM-generated reference is linked to its corresponding record in a bibliometric database, which enables downstream analysis. 
Bibliometric metadata is sourced from the database \textit{Dimensions}~\cite{hook2018dimensions}. 
References are initially matched based on their DOI; if the DOI is missing or malformed, a second attempt to match is made using the paper title. 
Author names, which are prone to format inconsistencies, are not used for matching. 
Once a reference is matched, we replace its original entry (author, publication year, etc.) with the corresponding record from Dimensions; unmatched references are removed from subsequent analyses.


\subsection{Data and models}

Full-text data are sourced from 1{,}746 main-track papers from ACL, EMNLP, and NAACL 2025 for which both \texttt{.tex} and \texttt{.bib} files are available from arXiv. 
Together, these papers contain 63{,}944 citation contexts corresponding to 132{,}913 citation slots, where each slot is a single cited work and a sentence citing multiple works contributes multiple slots. 
Table~\ref{tab:venue_distribution} summarizes the distribution of papers and matched citation contexts across venues. 

The experiment is repeated across six LLMs: GPT-5.1, Claude-3.5-Haiku, Gemini-2.0-Flash, DeepSeek-V3.2, Llama-4-Maverick, and Qwen2.5-72B-Instruct. For brevity, we refer to the last five as Claude-3.5, Gemini-2.0, DeepSeek, Llama-4, and Qwen-72B across figures. For citation-intent labeling, we use Gemini-3-Flash-Preview as the primary judge and replicate the results with DeepSeek-V4-Flash as a robustness check (Appendix~\ref{si_deepseek}).

\begin{table}[!htbp]
    \centering\small
    \begin{tabular}{lrr}
        \toprule
        Venue & \#Papers & Average \# contexts (SD)\\
        \midrule
        ACL   & $668$     & $38.9$ ($19.6$) \\
        EMNLP & $940$     & $36.0$ ($18.0$) \\
        NAACL & $138$     & $36.1$ ($15.2$) \\
        \midrule
        Total & $1{,}746$ & $37.1$ ($18.5$) \\
        \bottomrule
    \end{tabular}
    \caption{Corpus composition by venue. We study $1{,}746$ Main-track papers with available \texttt{.bib} and \texttt{.tex} on arXiv, and report the per-paper mean number of matched citation contexts with standard deviation in parentheses.}
    \label{tab:venue_distribution}
\end{table}

Of the $63{,}944$ human contexts with $132{,}913$ citation slots, $86.7\%$ are matched to a corresponding record in Dimensions (Table~\ref{tab:llm_general_count}). 
LLMs produce comparable volumes ($125$k--$132$k slots, Appendix~\ref{app:match}) but are matched at variable rates across models ($39.5$--$81.9\%$); since human references are matched at a high rate under the same pipeline, this gap is attributable to model behavior (e.g., hallucinating or malformed references; per-model disposition see Appendix~\ref{app:match}) rather than the matching procedure, revealing a basic departure from human citation practice. 
Given this variation, downstream analyses use the full matched set in the main text and are replicated on the intersection of contexts every model resolves (Appendix~\ref{si_intersect}), controlling for possible model effects.

\begin{table}[!htbp]
\centering\small
\begin{tabular}{lrrr}
\toprule
Source & Context & Citation & Matched (\%) \\
\midrule
Original (human)  & 63{,}944 & 132{,}913 & 115{,}278 (86.7) \\
DeepSeek-V3.2     & 61{,}869 & 125{,}810 & 103{,}044 (81.9) \\
GPT-5.1           & 58{,}019 & 125{,}176 & 88{,}219 (70.5) \\
Llama-4-Maverick  & 63{,}742 & 131{,}550 & 93{,}961 (71.4) \\
Gemini-2.0-Flash  & 63{,}883 & 131{,}910 & 62{,}744 (47.6) \\
Qwen2.5-72B       & 63{,}888 & 131{,}482 & 71{,}731 (54.6) \\
Claude-3.5-Haiku  & 63{,}664 & \multicolumn{1}{r}{131{,}416} & 51{,}878 (39.5) \\
\bottomrule
\end{tabular}
\caption{Corpus scope and Dimensions matching per model, across the 1{,}746
ACL/EMNLP/NAACL 2025 papers (DeepSeek-V3.2 covers 1{,}738). Contexts = distinct
(paper, context\_index); Citations = masked citation slots; Matched = slots matched to a Dimensions record.}
\label{tab:llm_general_count}
\end{table}

\section{LLMs cite less critically (RQ1)}\label{main:warming}

\begin{figure*}[!tbp]
    \centering
    \includegraphics[width=\linewidth]{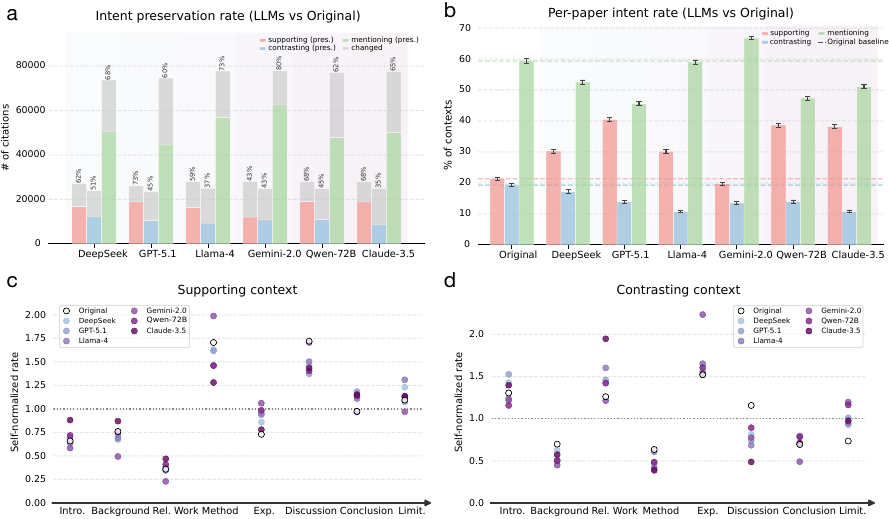}
    \caption{\textbf{LLMs warm the tone of citations, under-producing contrasting citations relative to humans.}
    \textbf{(a)} Intent preservation: of the citations the judge labeled on the human sentence (bar height $=$ \# citations), the colored portion keeps its label on the LLM-filled sentence (\% annotated). Contrasting is least preserved in every model ($34.6$--$50.6\%$).
    \textbf{(b)} Per-paper intent rate (each paper one observation, $N{=}1{,}746$; mean $\pm$ $95\%$ CI; dashed lines: human level). Models inflate supporting (five of six above human $21.3\%$, up to $40.6\%$) and suppress contrasting (all six below human $19.3\%$, down to $10.7\%$).
    \textbf{(c,d)} Self-normalized rate (per-section rate $\div$ cross-section mean; $1.0 =$ own average) for supporting (c) and contrasting (d) for selected sections. Within-source emphasis diverges: LLMs over-support in Experiments and Conclusion, under-support in Discussion; under-contrast in Background, Method, and especially Discussion, and over-contrast hugely in Limitations compared to humans.}
    \label{fig:motivation}
\end{figure*}

LLMs cite less critically than humans. 
Across the dataset, human citations split between 21\% supporting, 19\% contrasting, and 60\% mentioning, while LLM-filled sentences generated for each masked citation context produce \textbf{fewer \emph{contrasting}} citations, with a range of $10.7$--$17.2\%$ (Claude-3.5 and Llama-4 $11\%$, Gemini-2.0 $13\%$, GPT-5.1 $14\%$, Qwen-72B $14\%$, DeepSeek $17\%$). 
Five out of the six tested models also produce more supporting citations than the human-written baseline (GPT-5.1 $41\%$, Qwen-72B $39\%$, Claude-3.5 $38\%$, with Gemini-2.0 the lone exception at $20\%$; \autoref{fig:motivation}b). 
Under a second, independent judge, the contrasting share is even lower ($5$–$10\%$), so the deficit can not be attributed to an artifact of one judge.
The same trend is also observed when three human annotators annotate sentences manually, suggesting that LLMs genuinely generate fewer critical citations, rather than LLM judges failing to appropriately categorize sentences (Appendix~\ref{app:human_val}).

This shift can be generalized as a persistent directional ``warming'' of rhetorical intent.
The three citation sentence categories can be conceived as an ordered scale, progressing from contrasting, to mentioning, to supporting. 
Compared with the original human sentences, LLM-filled sentences persistently move up this scale across all models ($\Delta_{\text{cont}}=-2.4$ to $-8.6$\%).
That is, contrasting citation sentences written by humans are typically replaced by mentioning or supporting sentences, whereas human-written mentioning sentences are often replaced by supporting sentences. 
Indeed, the shift is asymmetric: when the human-written sentence is contrasting, the LLM-filled version is classified as \emph{supporting} in $12.5$--$31.1\%$ of cases (Claude-3.5 $31\%$, GPT-5.1 $29\%$, Qwen-72B $28\%$), whereas the reverse (supporting$\rightarrow$contrasting) is rare ($3.6$--$7.6\%$). 
That is, LLMs turn criticism into endorsement far more often than the opposite.

Critical contexts are the least likely to be preserved by the LLM. 
When the human-written citation sentence is contrasting, the LLM-filled sentence is assigned the same label only $34.6$--$50.6\%$ of the time (Claude-3.5 $35\%$, Llama-4 $38\%$, Gemini-2.0 $43\%$, GPT-5.1/Qwen-72B $45\%$, DeepSeek $51\%$), versus $60$--$80\%$ for mentioning (\autoref{fig:motivation}a).
In general, the agreement between the human-written and LLM-filled sentence classifications is at best fair (Cohen's $\kappa = 0.32$--$0.38$).

The self-reported intent of the LLM generating a replacement sentence is even ``warmer'' than when the same generated sentence is classified by an external LLM judge. 
Per their self-reported intent during generation, four of six models report $54$--$75\%$ supporting citation intent (Claude-3.5 $75\%$, GPT-5.1 $67\%$, Qwen-72B $59\%$, DeepSeek $54\%$), with Llama-4 ($47\%$) and Gemini-2.0 ($28\%$) lower but still above the human $21\%$, and almost no contrasting ($2.6$--$8.3\%$, except DeepSeek $17\%$; Figure~\ref{fig:si_motivation_shift}). 
The classifications by the LLM-as-judge temper this self-reported warmth. 
However, the warming effect persists (~\autoref{fig:motivation}b), and cannot merely be attributed to differences in model self-reports as opposed to externally-judged intent. 

Human-written and LLM-filled citations diverge in their rhetoric by the section of the paper, likely owing to differences in surrounding context. 
At the section level, reading each curve against its own self-normalized $1.0$ baseline (\autoref{fig:motivation}c, d; per-section rate $\div$ cross-section mean) clarifies exactly where humans and LLMs diverge by intent relative to their own baseline. 
For expository sections such as the Background and Related Work, LLM-filled sentences tend to have fewer contrasting citations compared to counterpart human-written sentences (Background $\times 0.45$--$0.60$ vs.\ human $\times 0.70$).
Method sections show a mild warming effect (LLMs $\times 0.40$--$0.62$ vs.\ human $\times 0.63$).
In the Experiments section of papers, LLM-filled citation sentences tend to more frequently be of both supporting and contrasting intent.
Humans use the Discussion section as a site of both support ($\times 1.72$) and contrast ($\times 1.16$), whereas every LLM generates contrasting citations far below this baseline ($\times 0.49$--$0.89$) and simultaneously places less weight on support ($\times 1.37$--$1.50$ vs.\ human $\times 1.72$). 
In sum, these findings suggest that LLMs produce more supporting and contrasting citations where claims can be grounded in concrete material (e.g., the experiments section), but fewer when claims must be carried by argument alone (e.g., the Discussion).



Despite this section-level variation, the overall pattern is consistent: LLMs systematically warm the tone of citations; they under-produce contrasting citations; they fail to preserve roughly half of genuinely critical contexts (rewriting them as supporting); they self-report as overwhelmingly supportive (Figure~\ref{fig:si_motivation_shift}). 
LLM-generated citation thus mutes scholarly disagreement, smoothing away the critical engagement that human citing encodes.

%
%
\section{Citation biases moderated by LLMs' intent (RQ2)}\label{main:bias}

\begin{figure*}[!tbp]
    \centering
    \includegraphics[width=\linewidth]{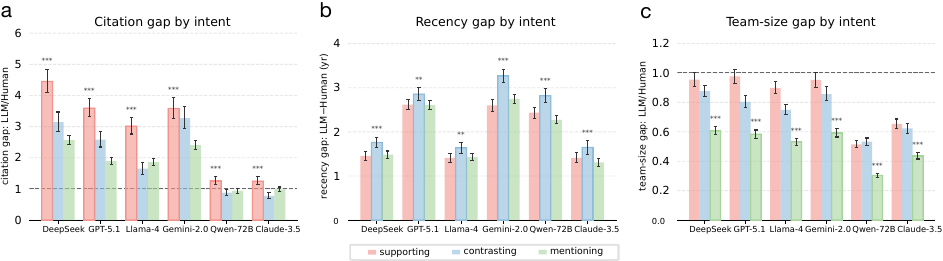}
    \caption{\textbf{LLM citation bias is amplified by rhetorical intent.} Human--LLM gap in three cited-paper attributes, by citation intent, aggregated at the \emph{paper} level (one observation per focal paper; 95\% CIs from between-paper variation). Citation count and team size are reported as \emph{geometric means} of per-paper ratios $\log(\text{LLM}/\text{human})$, back-transformed; recency is a mean per-paper difference. Intent is the judge's label on the LLM-filled (LLMs) or original (humans) sentence---intent-matched, not slot-matched. Dashed line marks parity ($1$ for ratios, $0$ for the difference). The most-deviant intent per model is outlined and starred (Welch $t$-test against the other two intents combined; $^{*}p{<}.05$, $^{**}p{<}.01$, $^{***}p{<}.001$). \textbf{(a)}~Citation count ratio ($>1$: cites more-cited work). \textbf{(b)}~Recency difference in years ($>0$: cites older work). \textbf{(c)}~Team-size ratio ($<1$: cites smaller teams). In every model the divergence peaks at a different intent per attribute: LLMs over-cite famous work most when \emph{supporting} (a), older work most when \emph{contrasting} (b), and under-cite large teams most when \emph{mentioning} (c).}
    \label{fig:llm_citation_bias}
\end{figure*}

We affirm prior findings that LLMs disproportionately cite highly-cited and smaller-team work~\cite{algaba_large_2025}. 
However, our own study also adds important nuance: these biases are not uniform, but moderated by citation intent. 
We examine how the human--LLM gap in characteristics of selected references varies across rhetorical intents, aggregating at the paper level: each focal paper contributes one observation per intent, so citation-heavy papers do not dominate.
Specifically, we compare the differences in citation count, team size, and recency between human-selected and LLM-filled references.
We report geometric-mean ratios for the heavy-tailed citation count and team size, and mean differences for recency, with confidence intervals reflecting between-paper variation.

The human baseline is itself strongly intent-dependent. 
When \emph{contrasting} a claim, humans tend to cite more recent ($2.25$\,yr) and less cited ($322$ citations) work.
When merely \emph{mentioning} background papers, humans pull in large, established consortium and benchmark papers ($25.6$ authors, $670$ citations, $3.07$\,yr). 
\emph{Supporting} citations sit in between ($16.5$ authors, $583$ citations, $2.93$\,yr). 
Citation intent is itself a strong signal of the type of paper an author is referencing.

The attributes of LLM-filled references diverge from human selections, though the magnitude of the difference varies by intent~(Figure~\ref{fig:llm_citation_bias}).
In terms of impact, LLMs cite papers $1.3$--$4.5\times$ as highly cited as humans do for the same citation type (DeepSeek $4.45\times$, GPT-5.1 $3.59\times$, Gemini-2.0 $3.58\times$).
The gap for contrasting citation is much smaller, and for some models even below parity (Qwen-72B $0.88\times$, Claude-3.5 $0.78\times$; Figure~\ref{fig:llm_citation_bias}a). 
In terms of paper recency, when generating a contrasting citation, LLMs cite work $1.6$--$3.3$ years older than humans, significantly above other intents in every model (Gemini-2.0 $+3.27$\,yr, GPT-5.1 $+2.86$\,yr, Qwen-72B $+2.82$\,yr; Figure~\ref{fig:llm_citation_bias}b), and this pattern persists among earlier cited works within every model's knowledge (Appendix~\ref{app:recency}). 
Regarding team size of cited papers, LLMs persistently cite papers with smaller teams across all intents. 
However, the gap is at its largest for mentioning citations, for which LLMs cite papers with teams only $0.30$--$0.61\times$ as large as the human baseline (Qwen-72B $0.30\times$, Llama-4 $0.53\times$), and lesser than for supporting or contrasting citations (Figure~\ref{fig:llm_citation_bias}c).
In each panel, the highlighted intent is significant against the other two intents combined in all six models (Welch $t$-test, $p<0.01$).

Two of the three peaks coincide with where human behavior is most distinctive: humans cite their most recent work when contrasting and papers with the largest teams when writing a mentioning citation; 
these cases are exactly where LLMs diverge most, toward older work with smaller teams. 
The citation-count peak is different.
When endorsing a claim, LLMs most strongly favor high-impact canonical papers, disproportionately citing them them when generating a supporting rather than a contrasting citation. 
These cases demonstrate that the divergence of human- and LLM-selected references is intent-dependent, robust across all six models, both judges (Appendix~\ref{si_deepseek}), and the shared-context intersection (Appendix~\ref{si_intersect}).

%
%
\section{Attenuated social proximity in LLM citation (RQ3)}

\begin{figure*}[t]
    \centering
    \includegraphics[width=\linewidth]{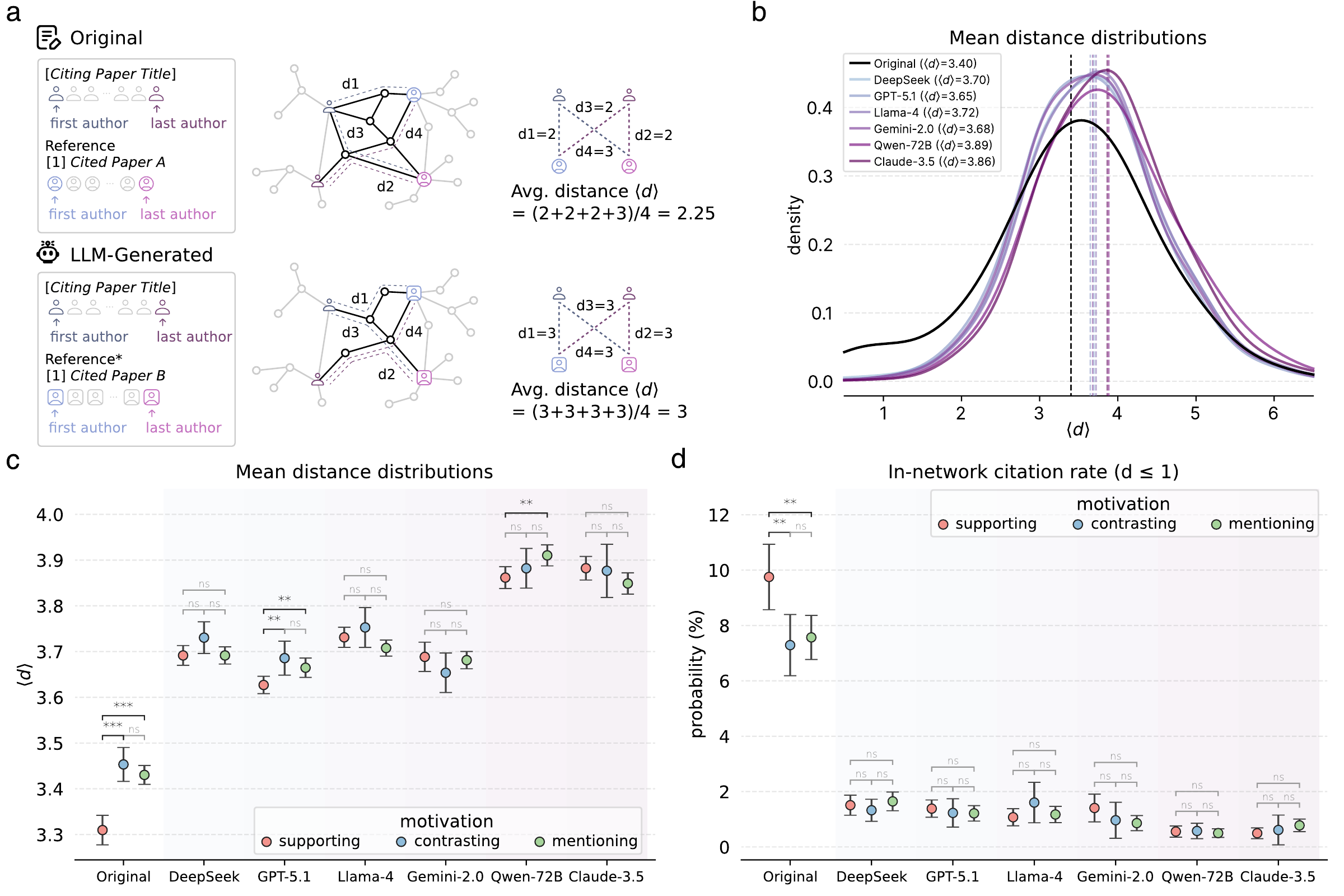}
\caption{\textbf{Author-dyad coauthorship distance separates human from LLM citation behavior.}
\textbf{(a)} For each masked slot, focal and cited first/last authors are resolved to researcher IDs and BFS shortest-path distance computed on a 20.3M-edge, 2.1M-researcher 2015--2024 coauthorship network; per-context mean $\langle d \rangle$ is the unit in (b)--(c).
\textbf{(b)} Density of $\langle d \rangle$: humans peak near $3.4$ with a close-circle shoulder; LLMs shift right ($3.65$--$3.89$).
\textbf{(c)} $\langle d \rangle$ by intent (mean $\pm$ 95\% CI; Mann--Whitney U). Humans show drastic gradient, supporting closer than contrasting/mentioning ($p<0.001$); LLMs cluster within $\pm 0.05$ hops.
\textbf{(d)} In-network citation rate ($d \leq 1$). Humans reach $7$--$10\%$, elevated for supporting ($9.8\%$ vs.\ $7.3\%$/$7.6\%$; $p<0.01$); every LLM stays below $1.6\%$ with no intent gradient.}
    \label{fig:network_proximity}
\end{figure*}

To understand whom LLMs and humans cite within their social networks, and how this differs by intent, we identify the first and last author of every focal and cited paper ($89$k$+$ unique authors across $104$k$+$ matched papers).
From these authors, we build a 10-yr (2015--2024) coauthorship network: an undirected edge links every pair of co-authors within any publication containing one of these authors (capped at a team size of 25). 
This publication-induced graph retains the 1-hop coauthor neighborhoods around our focal and cited authors, capturing the local collaboration structures most relevant to the observed citation pairs while providing a tractable approximation to the broader coauthorship network.
The resulting graph spans $2.1$M$+$ nodes and $20$M$+$ edges. 

For each citation slot, we compute shortest-path distances on the co-authorship graph using breadth-first search for four author-role dyads: focal-first$\leftrightarrow$cited-first, focal-first$\leftrightarrow$cited-last, focal-last$\leftrightarrow$cited-first, and focal-last$\leftrightarrow$cited-last. 
Here, $d=0$ denotes a self-citation, $d=1$ a direct collaborator, and larger values indicate increasingly distant colleagues; dyads in different connected components are coded as unreachable, and dyads with missing researcher identifiers are dropped. 
We summarize each citation by the average reachable distance $\langle d\rangle$ across these four dyads. 
Figure~\ref{fig:network_proximity}a illustrates this author-dyad distance computation. 
For intent-conditioned analyses, we further average citation-level distances within each (paper, context\_index) slot, so multi-citation contexts contribute one observation and over-cited slots do not dominate the distribution.

One constraint on this approach is the coverage of the coauthorship network.
Although $86.7\%$ of the $133$k$+$ human-written citations resolve to a publication record in Dimensions, only $26.5\%$ have at least one reachable author-role dyad in the 2015--2024 coauthorship graph; LLM citations yield reachable dyads for $14.5$--$25.5\%$ of slots, depending on the model. 
This yields $96$k$+$ human and $48$k--$85$k per-model reachable dyad observations.

Every LLM generates citations that are \textbf{farther} across the coauthorship network than in the human baseline. 
Figure~\ref{fig:network_proximity}b shows the distribution of per-context mean author-dyad distance across the seven sources. 
The human distribution peaks near 3 hops, reflecting a concentration in the near neighborhood of connections ($\langle d \rangle = 3.40$), while all six LLM distributions sit visibly to the right, with means ranging from $3.65$ (GPT-5.1) to $3.89$ (Qwen-72B). 
The shift is persistent across all six models, with no LLM recovering the human bias towards socially-proximate citations. 

Conditioning on intent (Figure~\ref{fig:network_proximity}c) sharpens this divergence, as LLMs vary their citation distance far less by rhetorical intent than do human authors.
For human citations, work referenced through a supporting citation is \emph{markedly more} socially proximate ($\langle d \rangle = 3.31$) than for contrasting ($3.45$) or mentioning ($3.43$) citations, a statistically significant difference ($p < 0.001$ for supporting vs.\ both other intents). 
None of the six LLMs reproduces these differences.
Their per-intent means cluster tightly within $\pm 0.05$ hops of each other, and any differences between intents are small and inconsistent across models (e.g., GPT-5.1 and Qwen-72B place supporting \emph{closer} than contrasting only marginally, while Claude-3.5 reverses it). 
LLMs treat the rhetorical role as orthogonal to who gets cited; humans do not. This human gradient holds within the same research field, while LLMs remain flat (Appendix~\ref{app:field}).

We further investigate socially-proximate citation practices (Figure~\ref{fig:network_proximity}d) to better understand how humans and LLMs diverge. 
We define the \emph{in-network citation rate} as the fraction of citation slots that are self-citations ($d=0$) or to a direct collaborator ($d=1$). 
Humans cite in-network ($d \leq 1$) at $7$--$10\%$ of slots, with the rate sharply elevated for supporting citations ($9.8\%$ supporting vs.\ $7.3\%$ contrasting, $p < 0.01$). 
For every LLM, the corresponding in-network citation rate is significantly lower, to $0.5$--$1.6\%$ regardless of intent, with self-citations ($d=0$) essentially absent. 
In other words, humans' tendency to cite socially-proximate research is not merely a mean shift in distance, but a tendency to disproportionately draw on their immediate social network, an effect that LLMs do not reproduce.

%
%
\section{Discussion}
We introduced a masked-citation framework to test whether LLMs cite like human scholars by intent. 
Unlike prior audits that treat citations as homogeneous retrieval outputs~\cite{tian_who_2024,he_who_2025,algaba_how_2026,algaba_large_2025}, our design aligns human and model-generated citation sentences, allowing direct comparison of how LLMs differ from humans in the selection and rhetoric of citation in the same context. 
We conducted experiments using six different LLMs representing the breadth of commercially available models, applying this task to 1.7k+ ACL/EMNLP/NAACL papers.
Three findings emerge from our experiments. 
First, LLMs cite \textbf{less critically}: they under-produce contrasting citations ($10.7$--$17.2\%$ vs.\ $19\%$ for humans) and lose the critical framing of about half the contexts humans wrote as contrasting, most often rewriting them as supporting.
Second, citation selection bias in LLMs is \textbf{moderated by intent}, and not merely a uniform bias toward popular papers. 
LLMs are more likely to generate citations to highly cited work in the case of a \emph{supporting} citation, older work most when \emph{contrasting}, and works with smaller teams of authors when \emph{mentioning}.
Third, LLMs \textbf{attenuate human biases related to social proximity} when selecting citations. 
That is, they do not inherit the human tendency to self-cite or cite close collaborators when \emph{supporting}; instead, LLMs generate socially distant citations across all intents.

\textit{Why} models differ from humans in their citation behavior remains an open question. 
The observed warming of citation intent may be related to the well-documented phenomenon of ``sycophancy''~\cite{Sharma2023TowardsUS}, a bias toward positivity and agreeableness in model outputs. 
Human-authored scientific writing is also carefully crafted to minimize emotion, and critique, when expressed, is delivered politely and cautiously~\cite{athar2014report}---tendencies that models may amplify. 
Concerning reference choice, humans draw on their social networks, for which LLMs have less knowledge and no equivalent to exposure through social proximity. 
What is harder to speculate on is why LLM citation selection varies by intent. 
One possible explanation lies in the training data.
Highly cited papers may more often appear in positive contexts, while more recent work appears in mixed or negative ones.
Future work is needed to explore mechanisms for observed biases, which will inform interventions and improvements in LLM citation.

Ongoing advancements in LLM tools may ameliorate observed biases. 
For example, LLM-based scientific assistants increasingly make use of agentic search to improve paper retrieval. 
Yet, such tools are not a panacea. 
The divergences we observe arise at the generation stage rather than from limited retrieval.
Giving GPT-5.1 live web search leaves the warming essentially unchanged (Appendix~\ref{app:rag}), consistent with evidence that LLM agents rely on parametric knowledge even when retrieval is available~\citep{fan_livebrowsecomp_2026}. 
Citation generation is not only a retrieval problem, but also a rhetorical and social process based on human judgment.

Our results provide important nuance to normative discourse surrounding the use of LLMs in science. 
On one hand, our findings demonstrate that LLMs, by drawing on less socially proximate papers, may expose authors to papers they may otherwise have never seen, offering a counter to increasingly narrow scholarly attention~\cite{varga2022narrowing, evans2008electronic}. 
On the other hand, LLMs smooth over critical rhetoric that a human might have written. 
Criticism is an essential component of science~\cite{decruz_2013_value, kitcher_advancement_1993, popper_logic_1959}; by warming rhetoric, LLMs may obscure critical divides in the scientific literature that are important for positioning viewpoints and findings. 

Citations have historically reflected careful choice on the part of their human authors and, although often biased~\cite{Leydesdorff2016}, were made with intent.
This intentionality made citations useful for positioning scientific works, tracing intellectual lineages, aiding literature search, and evaluating scholarly impact. 
Yet increasingly, citations are the product of machine generation rather than human deliberation, and these machines introduce biases of their own.
As LLMs become embedded in scientific writing, they should continue to be audited for potential bias, and researchers should be knowledgeable in their use and limits.
Awareness of these tendencies is a prerequisite for preserving critical engagement in scientific writing.

\newpage

\section*{Limitations}

\paragraph{LLM-as-judge models.} Intent labels come from an LLM judge rather than human annotators. The RQ1, RQ2, and RQ3 patterns replicate under a second judge (Appendix~\ref{si_deepseek}), but both judges may share biases inherited from LLM training distributions, for instance, reading characteristic LLM phrasing as more supportive than a human reviewer would. Our intent labels are therefore not validated against human judgment, and absolute rates (though not directional comparisons) should be interpreted with this in mind.

\paragraph{Matching coverage.} Dimensions match rates vary across models ($39.5$--$81.9\%$), so the main analyses use different supports per model. 
The shared-context intersection (Appendix~\ref{si_intersect}) reproduces all three intent-amplified patterns when every model contributes to the same contexts, ruling out coverage variation as the source of the bias.

\paragraph{Network coverage.} The 2015--2024 coauthorship network yields reachable dyads for only $26.5\%$ of human and $14.5$--$25.5\%$ of LLM citations. 
Authors with sparse publication footprints like early-career researchers, non-Western institutions, and industry practitioners are systematically underrepresented. 
The in-network gap we report therefore characterizes only research-active authors with traceable coauthorship, and may not generalize to citations of authors outside this subgraph.

\paragraph{Scope.} Our corpus is English-language ACL/EMNLP/NAACL main-track papers. 
Citation norms differ across disciplines, geographies, and languages, and the warming effect we observe, as well as biases toward popular papers, may be specific to this particular series of computer science conferences rather than a general phenomenon of LLM-assisted citation.

\section*{Ethical considerations}

Our study uses publicly available scholarly documents, papers from the main tracks of ACL, EMNLP, and NAACL 2025 (Dimensions.ai and arXiv), together with bibliometric and authorship metadata (Dimensions.ai) accessed under institutional subscription, from which we also construct our 2015--2024 coauthorship network. 
Resolving citing and cited authors involves personal data (names, affiliations, and career metadata) drawn entirely from these public publication records. 
We use it only to compute aggregate, model-comparative statistics, never to evaluate, rank, profile, or re-identify individual researchers, and we infer no protected or sensitive attributes; unresolved authors and unmatched citations are excluded rather than imputed. 
In line with the Dimensions terms, we do not redistribute raw records or the underlying network, and we will release only our code and aggregate, de-identified outputs.

We query the six LLMs through their providers' APIs in compliance with the respective terms of service, and corroborate our automatic intent labels with a second, independent judge. Our findings also point to concrete risks of LLM-assisted citation. 
Because LLM-generated citations are systematically biased toward highly cited and older work, and away from both critical (contrasting) citations and the citing author's own collaborators, uncritical reliance on such tools could entrench a canonical core, disadvantage recent, niche, or less-visible scholarship, and erode fairness in scholarly credit. 
By under-producing contrasting citations, these tools could also weaken critical engagement with prior work, and the masked-citation setup itself could be misused to fabricate or inflate citations. 
We report these findings to motivate auditing and human oversight of LLM-assisted citation, and we discourage such misuse.

\section*{Acknowledgments}
We are grateful for support from the NSF (Grant \#2219575).
We thank Digital Science for access to the Dimensions database, and the anonymous reviewers for their constructive feedback. We also thank the Network Science Institute at Northeastern University for computational resources.



\bibliography{references,references_dakota}

\newpage

\appendix

\section*{Appendix}

\section{Experimental design}

\subsection{Masking design}

Here we showcase our masking design for LLM citation generation and the LLM and system prompt (Figure~\ref{fig:mask-example}, Figure~\ref{fig:prompt}). We set the temperature to~$0$ for all models. 


\begin{figure}[!htbp]
\centering
\small
\textbf{(1) What the model sees} — one citation sentence is removed and replaced by \texttt{[CITE\_HERE]}:
\begin{lstlisting}[style=prompt]
Paper: "Leveraging Dual Process Theory in Language Agent Framework for Real-time Simultaneous Human-AI Collaboration"  (ACL 2025)
Section: Related Work
Theory of Mind (ToM) [CITE] has been introduced to enhance reasoning in human-AI collaborative scenarios [CITE]. [CITE_HERE] To adapt to humans, our framework integrates dual-process theory and ToM to support the entire process from perception to reasoning and decision-making.
Required citation count: 1
\end{lstlisting}
\textbf{(2) Human original} (held out; judge label: \emph{contrasting}):
\begin{lstlisting}[style=prompt]
"However, studies have pointed out that LLMs fail to achieve functional ToM (Riemer et al., 2024), where reasoning cannot be effectively implemented in decision-making processes."
\end{lstlisting}
\textbf{(3) Model reconstruction} (GPT-5.1; judge label: \emph{supporting}):
\begin{lstlisting}[style=prompt]
{
  "citation_count": 1,
  "recommended_papers": [
    {"title": "Theory of Mind May Have Spontaneously Emerged in Large Language Models",
     "authors": "Kosinski", "year": 2023,
     "venue": "arXiv", "doi": null}
  ],
  "citation_sentence": "Recent work has shown that large language models can exhibit emergent Theory of Mind-like capabilities (Kosinski, 2023).",
  "motivation": "supporting",
  "confidence": "low"
}
\end{lstlisting}
\caption{Masked-citation reconstruction (ACL 2025 paper). The model sees only the surrounding context with the citation sentence replaced by \texttt{[CITE\_HERE]} and the required citation count; it never sees the removed sentence or the cited work. Here the human sentence pushes back on prior work---``\emph{LLMs fail to achieve functional ToM}'' (\emph{contrasting})---while the model, on the same slot, recommends a different paper and reframes it as ``\emph{LLMs can exhibit emergent Theory of Mind}'' (\emph{supporting}): a representative instance of the tone warming we report.}
\label{fig:mask-example}
\end{figure}

\begin{figure}[!htbp]
\centering
\small
\textbf{System prompt}
\begin{lstlisting}[style=prompt]
You reconstruct removed citations from NLP papers. Given a paragraph where [CITE_HERE] replaces one removed citation-bearing sentence, infer the cited paper(s) and the replacement sentence using the surrounding context.

Return ONLY a valid JSON object, no markdown or text outside the JSON:
{
  "citation_count": <int>,
  "recommended_papers": [
    {"title": "...", "authors": "First, Second, ...", "year": <int>,
     "venue": "...", "doi": "..." or null}
  ],
  "citation_sentence": "<one sentence replacing [CITE_HERE]>",
  "motivation": "supporting" | "contrasting" | "mentioning",
  "confidence": "high" | "medium" | "low"
}

Rules:
- recommended_papers length must equal citation_count, which must equal
  the required count from the user prompt.
- Each item is one distinct work; no duplicates, no combining.
- citation_sentence is one complete sentence including the citation text.
- Prefer real, specific papers. If uncertain, give your best guess and
  set confidence to "low".

Motivation categories:
- supporting:  cited work provides evidence, methods, or findings aligned with the citing paper.
- contrasting: cited work is a competing approach, contradicting finding, or baseline the citing paper improves upon.
- mentioning:  cited work is referenced for background, definitions, or general acknowledgment.
\end{lstlisting}

\textbf{User prompt (template)}
\begin{lstlisting}[style=prompt]
Paper: "<paper title>"
Venue/year: <venue>, <year>
Section: <section>

<masked paragraph with [CITE_HERE], LaTeX clutter stripped>

Required citation count: <N>
\end{lstlisting}
\caption{Generation prompt. The system prompt fixes the JSON schema, the
one-distinct-work-per-citation rule, the required-count constraint, and the
three categories; the user prompt supplies the masked context and
the required number of citations for one slot.}
\label{fig:prompt}
\end{figure}

\subsection{LLM-as-judge for citation intent}

To compare the \emph{rhetorical stance} of human and model citations on equal footing, we score every citation slot with a single LLM judge (\texttt{Gemini-3-Flash-Preview}, temperature $0$). The judge labels each citation sentence as \emph{supporting}, \emph{contrasting}, or \emph{mentioning} with a confidence level, seeing only the sentence and its local context, never the cited paper's title and never the alternative version of the sentence (Figure~\ref{fig:judge-example}, Figure~\ref{fig:judge-prompt}).

\begin{figure}[!htbp]
\centering
\small
\textbf{(1) What the judge sees:} one citation sentence plus its surrounding context:
\begin{lstlisting}[style=prompt]
Paper: "Leveraging Dual Process Theory in Language Agent Framework for Real-time Simultaneous Human-AI Collaboration"  (ACL 2025)
Section: Related Work

Sentence before:
Theory of Mind (ToM) [CITE] has been introduced to enhance reasoning in human-AI collaborative scenarios [CITE].

Citation sentence:
<one of the two sentences below>

Sentence after:
To adapt to humans, our framework integrates dual-process theory and ToM to support the entire process from perception to reasoning and decision-making.
\end{lstlisting}

\textbf{(2a) Phase A — judging the HUMAN original} (judge label: \emph{contrasting}):
\begin{lstlisting}[style=prompt]
Citation sentence:
"However, studies have pointed out that LLMs fail to achieve functional ToM [CITE], where reasoning cannot be effectively implemented in decision-making processes."

-> {"motivation": "contrasting", "confidence": "high"}
\end{lstlisting}

\textbf{(2b) Phase B — judging the GPT-5.1 reconstruction} (judge label: \emph{supporting}):
\begin{lstlisting}[style=prompt]
Citation sentence:"Recent work has shown that large language models can exhibit emergent Theory of Mind-like capabilities (Kosinski, 2023)."

-> {"motivation": "supporting", "confidence": "high"}
\end{lstlisting}
\caption{LLM-as-judge classification. The judge receives only the citation sentence and the before/after window, cited-paper titles are withheld. The two phases score the human original (Phase A, run once per judge) and each model's reconstruction (Phase B, run per source$\times$judge) in independent calls, so the judge never anchors on the alternative version of the same slot. The judge confirms the tone warming: human sentence pushes back, while the model rewrites as supporting.}
\label{fig:judge-example}
\end{figure}

\begin{figure}[!htbp]
\centering
\small
\textbf{System prompt} (identical across Phase A and Phase B)
\begin{lstlisting}[style=prompt]
You classify citations in NLP papers as supporting, contrasting, or mentioning, based on a citation sentence and its surrounding context.

Categories:
- supporting: cited work provides evidence, methods, or findings aligned with the citing paper's claims or approach.
- contrasting: cited work is a competing approach, contradicting finding, or baseline the citing paper improves upon or disagrees with.
- mentioning: cited work is referenced for background, definitions, or general acknowledgment with no clear support or contrast.

Confidence:
- high: the surrounding text explicitly signals the relationship.
- medium: the relationship is strongly implied but not explicit.
- low: the relationship is ambiguous or inferable only with effort.

Return ONLY a valid JSON object, no markdown or text outside the JSON:
{
  "motivation": "supporting" | "contrasting" | "mentioning",
  "confidence": "high" | "medium" | "low"
}
\end{lstlisting}

\textbf{User prompt (template)}
\begin{lstlisting}[style=prompt]
Paper: "<paper title>"
Section: <section>

Sentence before:
<cleaned before-window, or "(beginning of paragraph)">

Citation sentence:
<original sentence (Phase A) OR model-filled sentence (Phase B), LaTeX clutter stripped>

Sentence after:
<cleaned after-window, or "(end of paragraph)">
\end{lstlisting}
\caption{Judge prompt. The system prompt fixes the JSON schema, the three
motivation categories, and the three confidence levels; the user prompt supplies
the citation sentence and a $\pm$500-character context window (sentence capped at
800 characters) with cited-paper titles deliberately omitted. The same prompt is
used to judge the human original and every model reconstruction, so any
difference in the label distribution is attributable to the sentence, not the
judging procedure.}
\label{fig:judge-prompt}
\end{figure}

To clarify how these definitions are applied, Table~\ref{tab:motivation_definition_example} provides representative examples of citation contexts categorized under each intent defined in Table~\ref{tab:motivation_definition}: \emph{Supporting}, where the reference provides evidence, methods, or findings aligned with the authors' claims or approach; \emph{Contrasting}, where the reference is a competing approach, contradicting finding, or baseline that is improved upon or disagreed with; and \emph{Mentioning}, where the reference is invoked for background, definitions, or general acknowledgment with no clear support or contrast.

\begin{table*}[t]
\centering\small
\begin{tabular}{p{0.18\linewidth} >{\raggedright\arraybackslash}p{0.74\linewidth}}
\toprule
Intent & Example \\
\midrule
Supporting \includegraphics[width=0.37cm]{figures/smile_salmon.png} &
\textbf{Paper:} ``Superpose Task-specific Features for Model Merging'' (arXiv:2502.10698) \\[2pt]
& \textbf{Section:} Introduction \\[2pt]
& \emph{Sentence before:} Representations in deep neural networks can be decomposed into combinations of feature vectors. \\[2pt]
& \emph{Citation sentence:} ``Recent works in mechanistic interpretability \colorbox{citehl}{(Bricken et al., 2023; Templeton et al., 2024)} validate the hypothesis and also reveal that these representations often contain features both related and unrelated to the model input.'' \\[2pt]
& \emph{Sentence after:} This phenomenon motivates our approach: linearly superposing features from individual models into the representation of the merged model can preserve task-specific capabilities. \\
\midrule
Contrasting \includegraphics[width=0.37cm]{figures/sad_blue.png} &
\textbf{Paper:} ``AttnComp: Attention-Guided Adaptive Context Compression for Retrieval-Augmented Generation'' (arXiv:2509.17486) \\[2pt]
& \textbf{Section:} Introduction \\[2pt]
& \emph{Sentence before:} While achieving high compression rates, they incur significant latency due to token-by-token decoding. \\[2pt]
& \emph{Citation sentence:} ``Extractive methods instead select relevant spans from the original content, offering greater efficiency \colorbox{citehl}{(Jiang et al., 2024; Hwang et al., 2024; Chirkova et al., 2025)}.'' \\[2pt]
& \emph{Sentence after:} However, current extractive methods typically only assess the relevance of individual sentence or document to the query, limiting their ability to integrate information across broader context. \\
\midrule
Mentioning \includegraphics[width=0.37cm]{figures/neutral_green.png} &
\textbf{Paper:} ``QAVA: Query-Agnostic Visual Attack to Large Vision-Language Models'' (arXiv:2504.11038) \\[2pt]
& \textbf{Section:} Related Work \\[2pt]
& \emph{Sentence before:} Initial research on adversarial attacks concentrated mainly on the visual modality, given its high-dimensional and continuous input space. \\[2pt]
& \emph{Citation sentence:} ``More recent studies have extended the attacks to discrete textual modalities \colorbox{citehl}{(Alzantot et al., 2018; Jia \& Liang, 2017; Wallace et al., 2019)}.'' \\[2pt]
& \emph{Sentence after:} Additionally, some research has focused on targeting the fusion of visual and textual modalities. \\
\bottomrule
\end{tabular}
\caption{Three examples of human-written citation sentences with their LLM-as-judge motivation label (Gemini-3-Flash, high confidence). Citation handles are highlighted within otherwise plainly-formatted sentences, reflecting that intent often depends on local context, not the citation sentence alone. Labels and definitions are inspired by \textit{Scite.ai} categories~\citep{nicholson_scite_2021}.}
\label{tab:motivation_definition_example}
\end{table*}

\section{Self-reported intent vs LLM-as-judge}\label{si:motivation-vs-judge}

\begin{figure*}[t]
    \centering
    \includegraphics[width=0.75\linewidth]{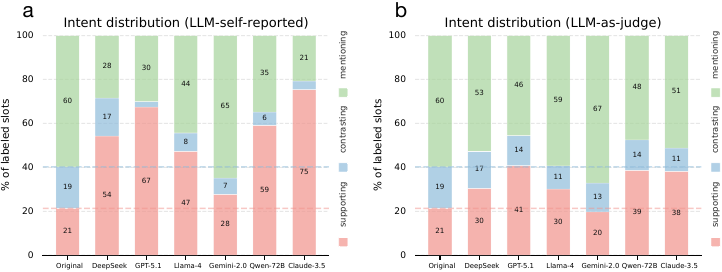}
\caption{\textbf{Intent distribution per source (\% of labeled slots; supporting / contrasting / mentioning).}
\textbf{a} LLM self-reported intent: original is identical (humans have no self-label, so we always read them through the judge), but each LLM bar uses the source LLM's own declared intent from during its generation of citations.
\textbf{b}: LLM-as-judge, the judge independently labels both the human original and each LLM's filled sentence; the same judging procedure is applied to every source.
Dashed lines mark the human supporting and supporting+contrasting baselines.}
    \label{fig:si_motivation_shift}
\end{figure*}

For every LLM the self-declared intent (LLM-self-reported) is warmer than the same model's writing as read by the judge (LLM-as-judge), which is already warmer than the human baseline (Figure~\ref{fig:si_motivation_shift}).
Self-supporting jumps to $28$--$75\%$ (Claude $75\%$, GPT-5.1 $67\%$, Qwen $59\%$, DeepSeek $54\%$, Llama-4 $47\%$, Gemini-2.0 $28\%$) versus the judge's $20$--$41\%$ on the same sentences, and self-contrasting collapses to $3$--$17\%$ versus $11$--$17\%$ under the judge.
Original is unchanged across the two panels because humans have no self-label.
Read together: every LLM both writes more supportive citations than humans and judges itself as more supportive than LLM-as-judge, the judge reads the LLM's text as a touch more critical than the LLM's own declared intent.

\section{Robustness check with DeepSeek as judge}\label{si_deepseek}

\begin{figure}[!htbp]
    \centering
    \includegraphics[width=\linewidth]{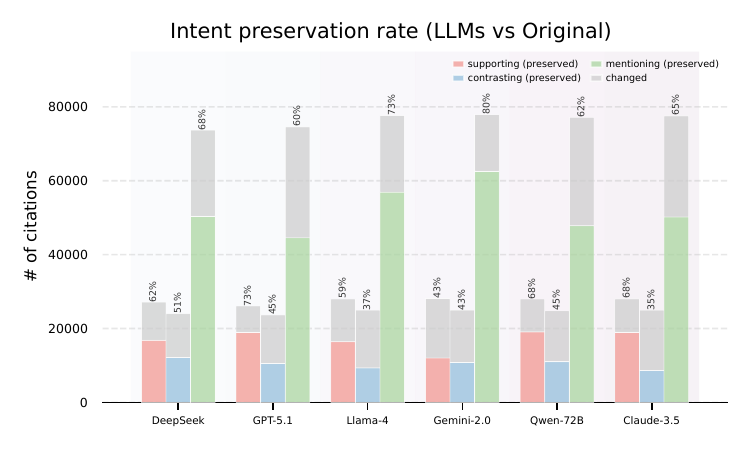}
    \caption{\textbf{Intent preservation under the DeepSeek-V4-Flash judge.}
Bar height shows the number of human citations labeled with each intent; the colored portion shows the share whose label is preserved in the LLM-filled sentence. As under the Gemini judge, \emph{contrasting} is least preserved for every model ($20.3$--$32.4\%$), while \emph{mentioning} is most preserved ($70.3$--$83.6\%$).}
\label{fig:ds_preservation}
    \label{fig:placeholder}
\end{figure}

Our main analyses use Gemini-3-Flash-Preview as the LLM-as-judge to label citation intent. To test whether the results depend on this particular judge, we repeat the intent-labeling step with DeepSeek-V4-Flash, using the same inputs, label definitions with temperature set to~$0$. This alternative judge labels both human original sentences and LLM-filled sentences under the \emph{same} supporting / contrasting / mentioning taxonomy defined in Table~\ref{tab:motivation_definition}.

The several results below show that some selected central patterns replicate under this alternative judge. LLM-generated citations remain less contrasting than human citations, citation biases continue to vary by rhetorical intent, and the human--LLM gap in coauthorship proximity remains qualitatively unchanged. This suggests that our findings are not driven by idiosyncrasies of a single judge model, but reflect robust differences between human and LLM citation behavior.

Beyond replicating the patterns, the two judges also agree at the label level on identical citation contexts (Table~\ref{tab:interjudge}): Gemini-3-Flash and DeepSeek-V4-Flash reach Cohen's $\kappa = 0.52$ on human originals and $0.45$ on the pooled LLM fills. Crucially, they rarely differ on the \emph{supporting} vs.\ \emph{contrasting} citations our claims rest on---one labels a citation \emph{supporting} while the other labels it \emph{contrasting} in only 2--9\% of cases---with disagreement falling almost entirely (92\%) on the neutral \emph{mentioning} boundary, which our findings do not depend on.

\begin{table}[!htbp]
\centering\small
\begin{tabular}{lrcc}
\toprule
Sentence set & $N$ & Cohen's $\kappa$ & Agreement \\
\midrule
Human original            & 57{,}958  & 0.52 & 72.6\% \\
LLM-filled & 335{,}004 & 0.45 & 68.9\% \\
\bottomrule
\end{tabular}
\caption{Agreement between the two LLM judges (Gemini-3-Flash vs.\ DeepSeek-V4-Flash) on identical citation contexts. Disagreement concentrates on the neutral \emph{mentioning} boundary rather than the \emph{supporting}/\emph{contrasting} citations our claims use.}
\label{tab:interjudge}
\end{table}

Repeating the per-slot intent preservation analysis from Figure~\ref{fig:motivation}a under the DeepSeek-V4-Flash judge reproduces a similar hierarchy (Figure~\ref{fig:ds_preservation}a).
Contrasting is the least-preserved intent in \emph{every} model ($20.3$--$32.4\%$), supporting sits in the middle ($25.3$--$41.9\%$), and mentioning is the most preserved ($70.3$--$83.6\%$).
Absolute preservation rates are lower than under the Gemini judge, contrasting drops to roughly half of what it was there ($34.6$--$50.6\% \to 20.3$--$32.4\%$), but the ordering is identical: when LLMs encounter a slot the judge reads as critical, they are the least likely to keep that critical framing on their own filled sentence, consistent with the warming trend reported in the main analysis.

\subsection{Bias amplification of LLM citations by intent}\label{si:deepseek_bias_amplification}

As a representative robustness check for RQ2, here we specifically revisit the recency result in Figure~\ref{fig:llm_citation_bias}b, where the human--LLM gap is largest for contrasting citations. Using the current DeepSeek-V4-Flash as a judge, we recover a similar pattern: for five of six models, LLMs cite older papers than humans most strongly in \emph{contrasting} contexts. The contrasting recency gap is significant for DeepSeek-V3.2 ($+1.63$ years, $p<0.01$), GPT-5.1 ($+2.80$ years, $p<0.05$), Gemini-2.0-Flash ($+3.08$ years, $p<0.001$), and Qwen2.5 72B Instruct ($+2.64$ years, $p<0.01$). Claude-3.5-Haiku shows the same direction but is not significant ($+1.50$ years), while Llama-4-Maverick shows similar gaps for contrasting and mentioning, with the mentioning gap reaching significance ($p<0.01$). Although the magnitudes are slightly smaller than under the Gemini judge, the same intent-level ranking is preserved, supporting the robustness of the contrasting-driven recency divergence.

\begin{figure}[!htbp]
    \centering
    \includegraphics[width=0.8\linewidth]{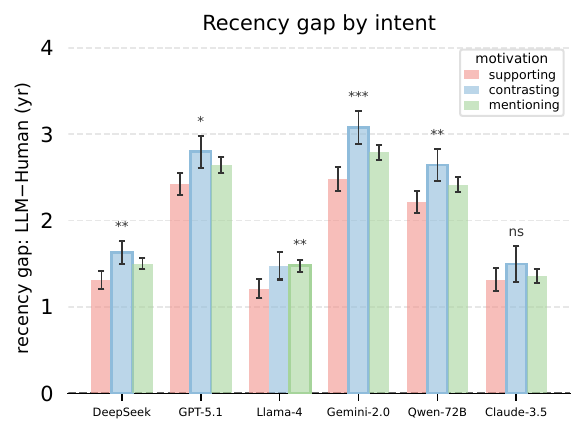}
    \caption{\textbf{Recency gap by intent under the DeepSeek-V4-Flash judge.}
    Human--LLM gap in cited-paper recency, aggregated at the paper level. Values show mean per-paper differences (LLM $-$ human, in years). The most-deviant intent per model is outlined and starred against the other two intents combined. As under the Gemini judge, the gap is largest for \emph{contrasting} citations in five of six models; Llama-4 splits between contrasting and mentioning.}
    \label{fig:ds_recency_gap}
\end{figure}

\subsection{Network proximity in LLM citations}\label{si:deepseek_network}

\begin{figure}[!htbp]
    \centering
    \includegraphics[width=\linewidth]{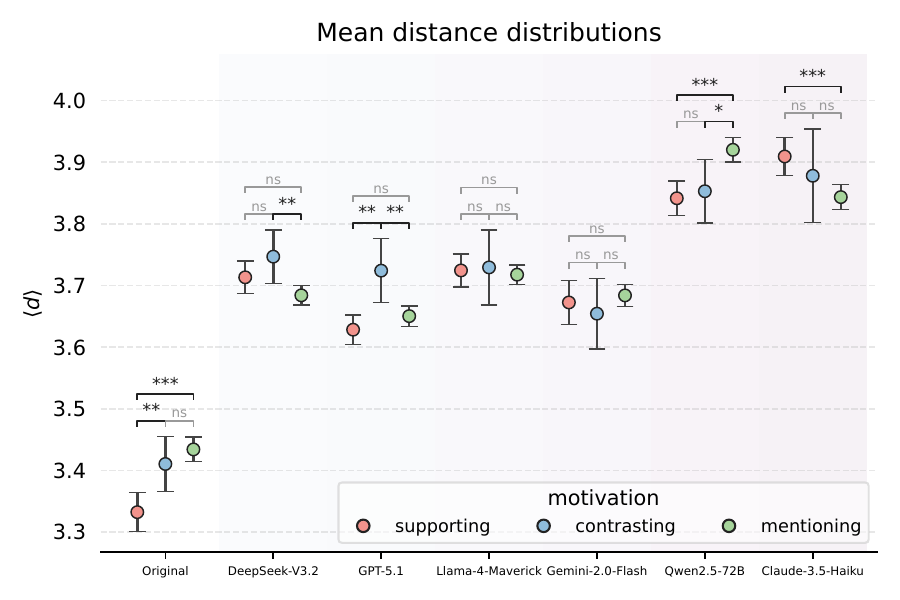}
    \caption{\textbf{Per context mean shortest path $\langle d \rangle$ by intent} (mean $\pm$ 95\% CI; Mann--Whitney U). Humans show a drastic gradient, supporting closer than contrasting/mentioning ($p<0.001$).}
    \label{fig:ds_distance}
\end{figure}

\begin{figure}[!htbp]
    \centering
    \includegraphics[width=\linewidth]{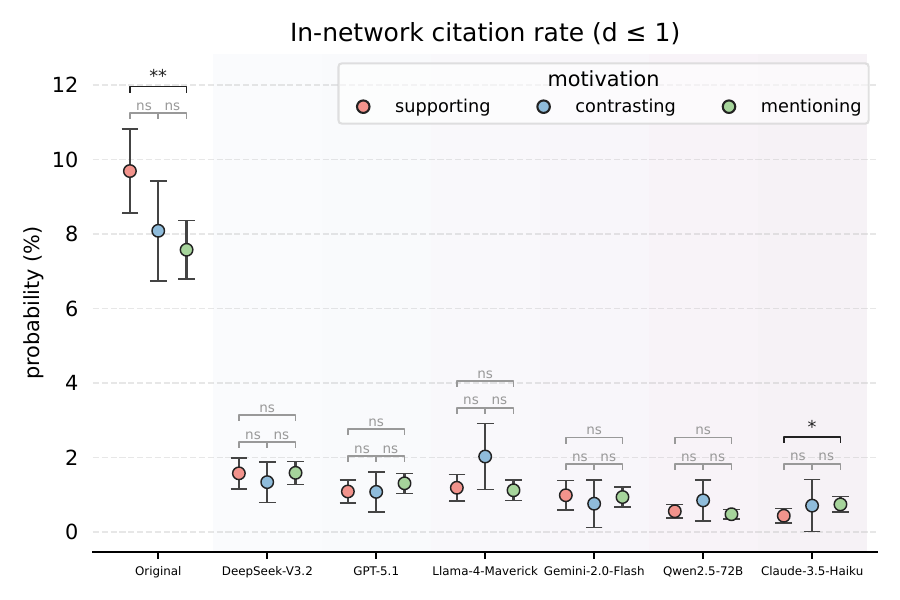}
    \caption{ \textbf{In-network citation rate ($d \leq 1$)}. Humans reach $7$--$11\%$, elevated for supporting; every LLM stays below $1.6\%$ with no/slight evident intent gradient.}
    \label{fig:ds_in_network}
\end{figure}

We re-ran the same network-proximity analysis with \emph{DeepSeek-V4-Flash} as the judge (Figures~\ref{fig:ds_distance}, \ref{fig:ds_in_network}); the pattern is unchanged.
Humans still cite markedly closer than every LLM ($\langle d \rangle = 3.33$ supporting vs.\ $3.65$--$3.92$ across LLMs), the human intent gradient survives (supporting vs.\ mentioning $p < 0.001$), and the in-network ($d \leq 1$) rate stays at $7.6$--$9.7\%$ for humans against $0.5$--$1.6\%$ for every LLM, elevated for human supporting and essentially flat across intent for every model.

The only small shifts are statistical: the human supporting-vs-contrasting effect is now marginal under this judge ($p \approx 0.07$ vs.\ $p < 0.01$ under Gemini), and Llama-4 / Gemini-2.0 / Claude-3.5 show no per-intent effect (vs.\ small but significant gaps under Gemini).
The headline is still that humans cite within their close coauthor neighborhood for \emph{supporting} citations, LLMs do not, holds verbatim across both judges.

\section{Human validation of intent labels}
\label{app:human_val}

Beyond our two LLM judges (Section~\ref{main:label_intent}, Appendix~\ref{si_deepseek}), we validate the intent labels against human annotators. Three annotators independently labeled a stratified sample of 90 sentences (30 per intent) under the same rubric, blind to the model labels and the cited paper. Taking their majority as ground truth, the primary judge (Gemini-3-Flash) matches on 73\% ($\kappa = 0.60$) and the second judge (DeepSeek-V4-Flash) on 60\% ($\kappa = 0.39$), both recovering \emph{contrasting} best (F1 = 0.73 and 0.72).

Per intent, the judge is most reliable exactly on the two non-neutral classes our claims turn on (Table~\ref{tab:human_val_perintent}): \emph{supporting} (F1 = 0.79) and \emph{contrasting} (F1 = 0.73, recall 0.80), which drive the ``warming'' result (fewer \emph{contrasting}, more \emph{supporting}; Section~\ref{main:warming}) and the intent-conditioned biases (Section~\ref{main:bias}). Disagreement falls mostly on the neutral \emph{mentioning} boundary, which our findings do not depend on.

\begin{table}[!htbp]
\centering\small
\begin{tabular}{lccc}
\toprule
Intent & Precision & Recall & F1 \\
\midrule
supporting            & 0.80 & 0.77 & 0.79 \\
\textbf{contrasting}  & 0.67 & \textbf{0.80} & \textbf{0.73} \\
mentioning            & 0.72 & 0.64 & 0.68 \\
\bottomrule
\end{tabular}
\caption{Per-intent precision/recall/F1 of the primary judge (Gemini-3-Flash) against the human majority on the 90-sentence validation sample. The judge is strongest on the two non-neutral classes (\emph{supporting}, \emph{contrasting}) that our claims rely on.}
\label{tab:human_val_perintent}
\end{table}

We also test whether the LLM judge simply reads LLM-written sentences as warmer. On the same 90 slots, the three annotators also labeled GPT-5.1's reconstruction of each citation. The warming we report (Section~\ref{main:warming}) reappears in the \emph{human} labels---\emph{contrasting} falls and \emph{supporting} rises for \emph{every} annotator, the same direction both LLM judges show (Table~\ref{tab:human_val_warming}). The judge also agrees with the human majority on these LLM-written sentences about as well as on human ones ($\kappa = 0.51$ vs.\ $0.60$), with no drop in human--human agreement on the LLM text. The warming is thus a property of the regenerated citations, captured by humans and both judges alike, not a judge self-preference for LLM-generated text.

\begin{table}[!htbp]
\centering\small
\begin{tabular}{lcc}
\toprule
Rater & contrasting & supporting \\
\midrule
Human majority vote  & 26\% $\rightarrow$ 18\% & 37\% $\rightarrow$ 55\% \\
Human annotator 1    & 26\% $\rightarrow$ 16\% & 23\% $\rightarrow$ 55\% \\
Human annotator 2    & 29\% $\rightarrow$ 22\% & 39\% $\rightarrow$ 49\% \\
Human annotator 3    & 26\% $\rightarrow$ 21\% & 42\% $\rightarrow$ 56\% \\
LLM judge (Gemini)   & 35\% $\rightarrow$ 19\% & 34\% $\rightarrow$ 56\% \\
LLM judge (DeepSeek) & 27\% $\rightarrow$ \phantom{0}9\% & 22\% $\rightarrow$ 39\% \\
\bottomrule
\end{tabular}
\caption{Intent rates on the original (human) vs.\ GPT-5.1-filled sentence, by rater, over the 77 of 90 slots for which GPT-5.1 returned a citation sentence (majority vote = 2 of 3 annotators). }
\label{tab:human_val_warming}
\end{table}

\section{Citation disposition and unmatched references}
\label{app:match}

All six models fill the same 63{,}944 citation contexts (132{,}913 requested citations). Produced counts differ only by each model's \emph{format-error rate}, the share of malformed or unparseable responses we drop; the remaining citations are matched to Dimensions (Table~\ref{tab:disposition}). Format errors are limited (0.8--5.8\%), and the variation in match rate (39.5--81.9\%) lies in the unmatched, well-formed bucket, which all downstream analyses exclude.

\begin{table}[!htbp]
\centering\small
\begin{tabular}{lrrr}
\toprule
Model & citations & format-error & matched\\
\midrule
DeepSeek-V3.2    & 125{,}810 & 5.3\% & 81.9\% \\
Llama-4-Maverick & 131{,}550 & 1.0\% & 71.4\% \\
GPT-5.1          & 125{,}176 & 5.8\% & 70.5\% \\
Qwen2.5-72B      & 131{,}482 & 1.1\% & 54.6\% \\
Gemini-2.0-Flash & 131{,}910 & 0.8\% & 47.6\% \\
Claude-3.5-Haiku & 131{,}416 & 1.1\% & 39.5\% \\
\bottomrule
\end{tabular}
\caption{Per-model citation disposition. Format errors are small, and the match-rate spread lies in the unmatched (well-formed) bucket.}
\label{tab:disposition}
\end{table}

The unmatched bucket is not used downstream, but to characterize the unmatched bucket, we audited 100 randomly sampled unmatched titles each from GPT-5.1 (high match rate) and Claude-3.5-Haiku (lowest), using Claude Opus 4.8 with web search, into three categories: a precise match we missed, a real work under an inaccurate title, or a fabricated reference (Table~\ref{tab:audit}). Most are hallucinated (B+C): 79\% for GPT-5.1 and 97\% for Claude-3.5-Haiku, and the higher-matching model hallucinates less.

\begin{table}[!htbp]
\centering\small
\begin{tabular}{lcc}
\toprule
Unmatched title & GPT-5.1 & Claude-3.5-Haiku \\
\midrule
A. precise match            & 21\% & \phantom{0}3\% \\
B. real work, garbled title & 51\% & 24\% \\
C. fabricated               & 28\% & 73\% \\
\bottomrule
\end{tabular}
\caption{Audit of 100 unmatched titles per model. Most are hallucinations (B+C): 79\% for GPT-5.1, 97\% for Claude-3.5-Haiku.}
\label{tab:audit}
\end{table}

\section{Robustness to context window}
\label{app:context}

\begin{table}[!htbp]
\centering\small
\setlength{\tabcolsep}{4pt}
\begin{tabular}{lccc}
\toprule
Context window & Sup.~\includegraphics[width=0.28cm]{figures/smile_salmon.png} & Con.~\includegraphics[width=0.28cm]{figures/sad_blue.png} & Men.~\includegraphics[width=0.28cm]{figures/neutral_green.png} \\
\midrule
$\pm 1$-sentence (baseline)    & 40.1 & 10.0 & 49.8 \\
+ full paragraph ($2.6\times$) & 45.1 & 10.4 & 44.5 \\
+ paragraph \& abstract        & 43.1 & 10.9 & 46.0 \\
\bottomrule
\end{tabular}
\caption{GPT-5.1 intent distribution (\%) as the input context widens (stratified 600-slot subset, 200 per intent; judge: DeepSeek-V4-Flash). Sup./Con./Men.\ = supporting/contrasting/mentioning. \emph{Contrasting} stays low and \emph{supporting} high across all levels.}
\label{tab:context}
\end{table}

To test whether our $\pm 1$-sentence window shapes the intent distribution, we re-ran GPT-5.1 on a stratified subset of 600 slots (200 per intent) with progressively more of the manuscript, holding the paper title, venue, section, and required citation count fixed: the $\pm 1$-sentence window (baseline), the full paragraph (about $2.6\times$ the context), and the paragraph plus the citing paper's abstract. Intent is labeled by DeepSeek-V4-Flash. The distribution barely shifts across conditions (Table~\ref{tab:context}): \emph{contrasting} stays at 10--11\% and \emph{supporting} remains high, and adding context does not move the model toward the human distribution. The intent distribution is thus robust to the amount of context provided.

\section{Robustness of the recency gradient}
\label{app:recency}

The tendency to cite older work is not merely limited access to recent papers: it persists even among cited works from before 2024, which are well within every model's knowledge. Restricting both human and LLM citations to these earlier works and recomputing the per-paper, intent-matched gap, the intent gradient holds: \emph{contrasting} remains the peak-recency intent in every model and stays significant throughout (Table~\ref{tab:recency}).

\begin{table}[!htbp]
\centering\small
\begin{tabular}{lcc}
\toprule
Model & $\Delta$ (yr, cited $<$ 2024) & $p$ \\
\midrule
DeepSeek-V3.2    & +0.8 & $<0.001$ \\
GPT-5.1          & +0.4 & $<0.01$ \\
Llama-4-Maverick & +0.6 & $<0.001$ \\
Gemini-2.0-Flash & +1.0 & $<0.001$ \\
Qwen2.5-72B      & +1.0 & $<0.001$ \\
Claude-3.5-Haiku & +0.7 & $<0.001$ \\
\bottomrule
\end{tabular}
\caption{Recency gradient on cited works from before 2024. $\Delta > 0$ means \emph{contrasting} citations show a larger human--LLM recency gap than the other intents; the gradient holds and stays significant in every model.}
\label{tab:recency}
\end{table}

\section{Robustness check with slot intersections}\label{si_intersect}

\begin{figure*}[t]
    \centering
    \includegraphics[width=\linewidth]{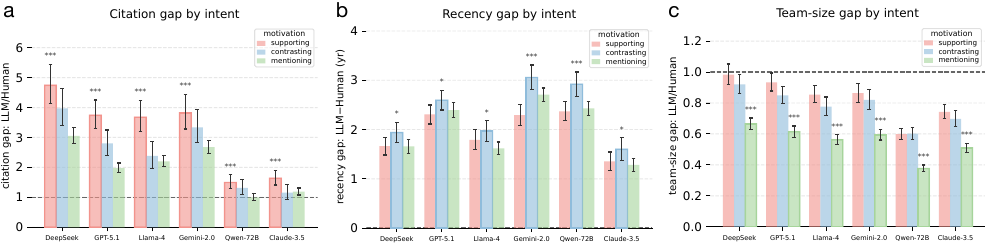}
\caption{\textbf{Intent-amplified citation bias replicates on the shared-context intersection.} Recomputed on the $12{,}556$ contexts ($1{,}695$ of $1{,}746$ papers) where every source---human original and all six LLMs, has $\geq 1$ Dimensions-matched cite. Per-paper aggregation: geometric-mean ratios for \textbf{a} citation count and \textbf{c} team size (LLM/Human; $>1$ = more-cited, $<1$ = smaller teams), mean year difference for \textbf{b} recency ($>0$ = older); $95\%$ CIs from between-paper variation; outlined intent = most-deviant per model (Welch $t$-test vs.\ the other two combined; $^{*}p{<}.05$, $^{**}p{<}.01$, $^{***}p{<}.001$). The headline pattern reproduces: citation gap peaks at \emph{supporting} ($3.7$--$4.7\times$ across top models), recency at \emph{contrasting} ($+2.6$ to $+3.1$\,yr), and team size at \emph{mentioning} ($0.38$--$0.56\times$). The intent-amplified bias is not an artifact of per-LLM matching coverage.}
    \label{fig:llm_citation_bias_intersect}
\end{figure*}

A natural concern is that the per-paper attribute gaps in Figure~\ref{fig:llm_citation_bias} are driven by \emph{which} contexts each LLM happened to land matches on rather than by the citations themselves, since Dimensions match rates vary widely across models ($39.5$--$81.9\%$). To rule this out, we recompute the same per-paper, per-intent gaps on the intersection of contexts where every source, the human original plus all six LLMs, has at least one Dimensions-matched cite, so every model is evaluated on identical support. This intersection retains $12{,}556$ contexts across $1{,}695$ of $1{,}746$ focal papers ($97\%$), trading per-paper context depth for an apples-to-apples comparison without losing the underlying population, though confidence intervals widen accordingly at the smaller sample size. 

The intent-amplified pattern from Figure~\ref{fig:llm_citation_bias_intersect} reproduces in both direction and magnitude: the citation-count gap peaks at \emph{supporting} for every model (DeepSeek $4.74\times$, Gemini-2.0 $3.82\times$, GPT-5.1 $3.74\times$, Llama-4 $3.68\times$), the recency gap peaks at \emph{contrasting} (Gemini-2.0 $+3.06$\,yr, Qwen $+2.92$\,yr, GPT-5.1 $+2.60$\,yr), and the team-size gap peaks at \emph{mentioning} (Qwen $0.38\times$, Llama-4 $0.56\times$). Because every model now contributes to the same context set, this design is more conservative than the main analysis: any remaining gap cannot be attributed to differential coverage. The persistence of the intent-amplified pattern therefore confirms that it reflects what the models choose to cite, not which slots they happen to resolve.

\section{Warming under retrieval}
\label{app:rag}

Our main setup asks models to cite from parametric knowledge. To test whether the warming depends on this, here we add a retrieval-augmented condition: GPT-5.1 answers each slot with live web search enabled (tool use), retrieving candidates as a RAG system would, on the same stratified subset ($n=599$; judge: DeepSeek-V4-Flash). The intent distribution stays essentially the same (Table~\ref{tab:rag}): even with live search, the model under-produces \emph{contrasting} citations (about 10\% vs.\ 33\% for humans), so the warming persists under retrieval.

\begin{table}[!htbp]
\centering\small
\setlength{\tabcolsep}{4pt}
\begin{tabular}{lccc}
\toprule
Intent source & Sup. & Con. & Men. \\
\midrule
human (original)          & 33.2 & 33.4 & 33.4 \\
GPT-5.1, parametric       & 39.9 & 10.9 & 49.2 \\
GPT-5.1, + live web search & 44.4 & 10.2 & 45.4 \\
\bottomrule
\end{tabular}
\caption{Intent distribution (\%) with and without live web search, on the balanced subset (200 per intent, so human rates are near-equal; $n=599$; judge: DeepSeek-V4-Flash). \emph{Contrasting} stays near 10\% under retrieval. Sup./Con./Men.\ = supporting/contrasting/mentioning.}
\label{tab:rag}
\end{table}

\section{Proximity gradient within research fields}
\label{app:field}

We test whether the intent-conditioned proximity gradient reflects field structure rather than social proximity. Using the paper-level Fields of Research (FoR) that Dimensions assigns to every publication (ANZSRC 2020; L1 = 2-digit division, L2 = 4-digit group), we label a citation \emph{within-field} if the citing and cited papers share any FoR code, and re-examine the citation-dyad distances for GPT-5.1 (Table~\ref{tab:field}). Holding field constant, humans still cite closer in every stratum (a gap of $+0.31$ to $+0.39$ hops, all $p < 10^{-6}$), and the supporting-closer gradient persists among same-field citations (human \emph{supporting} 3.00 vs.\ 3.40 for the other intents at L2), while GPT-5.1 stays flat. The proximity pattern thus holds within fields, not only across them. It does not distinguish social proximity from legitimate expertise among nearby researchers, which we note in the Limitations.

\begin{table}[!htbp]
\centering\small
\setlength{\tabcolsep}{4pt}
\begin{tabular}{llccc}
\toprule
FoR & field stratum & human $\langle d\rangle$ & GPT-5.1 $\langle d\rangle$ & gap \\
\midrule
L1 & within-field  & 3.31 & 3.62 & $+0.31$ \\
L1 & between-field & 3.31 & 3.70 & $+0.39$ \\
L2 & within-field  & 3.30 & 3.62 & $+0.32$ \\
L2 & between-field & 3.31 & 3.66 & $+0.35$ \\
\bottomrule
\end{tabular}
\caption{Mean coauthorship distance $\langle d\rangle$ by field stratum for GPT-5.1 vs.\ humans. Humans cite closer within \emph{and} between fields (all gaps $p < 10^{-6}$), so the proximity gap is not explained by field structure.}
\label{tab:field}
\end{table}

\section{AI assistance disclosure}
We used AI assistants, specifically ChatGPT and Claude, for language editing, polishing, and wording suggestions. All substantive research ideas, experimental design, analyses, claims, and final writing decisions were made and verified by the authors. The authors take full responsibility for the content of the paper.

\end{document}